%% file: CPP2016proc.tex
\documentclass[a4paper]{jpconf}
\usepackage{graphicx}
\usepackage{cite}

\pdfoutput=1

\def\be{\begin{equation}}
\def\ee{\end{equation}}
\def\bea{\begin{eqnarray}}
\def\eea{\end{eqnarray}}
\def\nn{\nonumber}

\newcommand{\secdec}{{\textsc{SecDec}\,\,}}
\newcommand{\pysecdec}{py{\textsc{SecDec}\,}}

\newcommand{\idel}{i\,\delta}
\newcommand{\eps}{\epsilon}
\newcommand{\rd}{{\mathrm{d}}}

\begin{document}


\title{Multi-loop calculations: numerical methods and applications}

\author{S.~Borowka$^1$,
G.~Heinrich$^{2,*}$,
S.~Jahn$^2$,
S.~P.~Jones$^2$,
M.~Kerner$^2$,
J.~Schlenk$^3$}
\footnotetext{$^*$Speaker. To appear in the proceedings
  of the 4th Computational Particle Physics Workshop, October 2016, Hayama, Japan.}
\address{$^1$ Theoretical Physics Department, CERN, Geneva, Switzerland}
\address{$^2$ Max Planck Institute for Physics, F\"ohringer Ring 6, 80805 Munich, Germany}
\address{$^3$ IPPP, University of Durham, Durham DH1 3LE, UK}

\begin{abstract}
We briefly review numerical methods for calculations beyond one loop and then describe new developments within the 
method of sector decomposition in more detail. We also discuss applications to two-loop integrals involving several mass scales.
\end{abstract}

\section{Introduction}

Precision calculations are of primary importance to scrutinise the Standard Model (SM) of particle physics and in particular the Higgs sector, 
where experiments moved from the discovery phase to the  phase of
precision measurements of the Higgs properties.
Small deviations from  the expected values  may be our only hints to physics beyond the SM for some time, 
and therefore precise theoretical predictions are mandatory.

 In the last decade, predictions at next-to-leading
 order (NLO) in perturbation theory in the strong coupling constant
 $\alpha_s$ got a large boost due to advances in calculational methods,
 and, together with NLO matching to parton shower Monte Carlo programs, 
 became the state of the art to describe the data. 
 However, for the phases II and III of the LHC, and even more so at future colliders,
the situation is drastically different: the experimental precision for 
many important SM processes already has reached a level where NLO QCD predictions fall short.
Therefore, a lot of effort has been
 spent in the past years to come up with corrections going beyond NLO
 QCD, 
 ideally not only for total cross sections, but also for differential
 distributions. 

\vspace*{3mm}

A measure of complexity for the calculation of higher order corrections in perturbation theory involves the number of loops in the virtual amplitude,
the number of scales (Mandelstam invariants, masses) and the number of external legs.
While the problem of infrared subtractions is more severe the more massless particles are involved, the difficulty to obtain analytic expressions for 
master integrals at two loops and beyond increases rapidly as the
number of mass scales grows.
Therefore numerical methods to calculate loop integrals seem
particularly well suited for integrals with several mass scales.

\vspace*{3mm}

For  processes involving only massless particles,  virtual two-loop 4-point
amplitudes have been calculated about 15 years
ago~\cite{Glover:2001af,Garland:2001tf,Bern:2002tk,Moch:2002hm,Binoth:2002xg}.
After that, the main bottleneck to be overcome to achieve NNLO
predictions for  processes involving massless two-loop 4-point
amplitudes 
was the lack of an efficient subtraction scheme for the infrared singularities
occurring in the {\em real radiation part}, where up to two particles can
become unresolved. Filling this gap has been a very active field of
research in the past years. Various methods have been devised  and
are still under active development; they will be listed briefly below. 

\vspace*{3mm}

The next big step in the field of NNLO QCD corrections for $2\to 2$
scattering processes was the 
availability of results for processes involving massive particles.
Here the main problem currently resides in the {\em virtual two-loop part} of the calculation.
Two problems are hampering progress here:  (a) the {\em reduction} of the two-loop amplitudes to a minimal set 
of ``master integrals" times coefficients gets increasingly complicated as the number of mass scales grows, and 
(b) the {\em  analytic calculation} of the master integrals is extremely difficult, 
entering unexplored territory in terms of mathematical functions to express the occurring parameter integrals.
Only very recently, analytic representations of two-loop integrals and amplitudes for $2\to 2$
scattering processes involving massive particles
became available, see e.g.~\cite{Henn:2013pwa,Henn:2014lfa,Caola:2014lpa,Gehrmann:2015ora,vonManteuffel:2015msa,Caola:2015ila,Bonciani:2016ypc,Bonciani:2016qxi,Primo:2016ebd}.


\section{Methods and tools for two-loop calculations and beyond} 

The steps to perform for the calculation of a (multi-)loop amplitude 
can be roughly divided into four stages: (1) generation of algebraic
expressions for the amplitude, (2) reduction of the amplitude to a set
of ``master integrals'' times coefficients, (3) isolation  of the
ultraviolet and infrared poles and  (4) evaluation of the master
integrals and combination with the coefficients to obtain the amplitude. 
To calculate a full cross section, loop amplitudes and real radiation
contributions need to be combined, which requires a suitable
scheme for the isolation of infrared-divergent real radiation, which is
highly non-trivial beyond one loop. It also requires the construction of a
stable and fast Monte Carlo program to perform the
phase space integration.

It should be mentioned that step (2) above is not
mandatory. Reducing the set of integrals to a minimal ``basis set'' is
usually beneficial,  to reduce the number of integrals to calculate
and to avoid large cancellations between linearly dependent
integrals. However, it is also possible to evaluate the occurring
integrals without reduction in a numerical approach, see e.g.~\cite{Spira:1995rr,Becker:2012bi,Borowka:2016ehy}.
Further,  there are methods which aim to avoid the problems with IR singularities related to the split
into real and virtual contributions  by not performing
such a partition at all~\cite{Soper:1999xk,Forde:2003jt,Catani:2008xa,Sborlini:2016hat}.

In Table~\ref{tab:loopamp} we give a list of some publicly available multi-purpose tools which have been
developed to perform the specific tasks described above, focusing on the numerical evaluation of the loop integrals.
Certainly this list is incomplete and omits a multitude of codes which may be more efficient, but
are tailored to more specific classes of integrals or amplitudes.
Efforts towards the development of a package that can provide all the
steps listed in Table~\ref{tab:loopamp} by combining {\sc Qgraf}~\cite{Nogueira:1991ex}, {\tt
  FORM}~\cite{Vermaseren:2000nd,Kuipers:2013pba}, {\sc Reduze}~\cite{Studerus:2009ye,vonManteuffel:2012np} and \pysecdec\cite{Borowka:2017idc} are described in~\cite{Jones:2016bci}.

Table \ref{tab:realsub} shows some of the subtraction schemes for
infrared divergent real radiation at NNLO.

\begin{center}
\begin{table}[h]
\caption{\label{tab:loopamp}Public tools for various steps of loop
  amplitude calculations beyond one loop.}
\centering
\begin{tabular}{@{}*{7}{l}}
\br
Step to be performed& available public tools\\
\mr
Diagram generation&{\sc Qgraf}~\cite{Nogueira:1991ex}, {\sc
  FeynArts/FormCalc}~\cite{Hahn:2000kx,Hahn:2016ebn}\\
Amplitude manipulations& {\sc Diana}~\cite{Tentyukov:1999is}, {\sc
  FeynCalc}~\cite{Shtabovenko:2016sxi,Shtabovenko:2016whf}\\
Reduction &{\sc
  Reduze}~\cite{Studerus:2009ye,vonManteuffel:2012np}, {\sc
  Fire}~\cite{Smirnov:2008iw,Smirnov:2014hma}, {\sc
  LiteRed}~\cite{Lee:2012cn,Lee:2013mka}, {\sc Air}~\cite{Anastasiou:2004vj}\\
Numerical evaluation &sector\_decomposition~\cite{Bogner:2007cr},
{\sc SecDec}~\cite{Borowka:2015mxa,Borowka:2017idc}, {\sc Fiesta}4~\cite{Smirnov:2015mct},\\
of the loop integrals&{\sc Nicodemos}~\cite{Freitas:2012iu}, {\sc Ambre/MBnumerics}~\cite{Gluza:2010rn,Dubovyk:2016ocz}\\
\br
\end{tabular}
\end{table}
\end{center}
Concerning the reduction, we only listed the publicly available tools
which, based on the integration-by-parts (IBP) method~\cite{Laporta:2001dd}, can be
used within a completely automated setup.
Ideas how  to reduce  the computational complexity of IBP algorithms
can be found in \cite{vonManteuffel:2014ixa}.
A fully automated system for amplitude generation and evaluation is also given by
the {\sc Grace} system~\cite{Belanger:2003sd,Fujimoto:2011xea,Khiem:2014dka}.

Novel reduction methods (see e.g.~\cite{Kosower:2011ty,Mastrolia:2012wf,Badger:2012dp,Mastrolia:2013kca,Larsen:2015ped,Ita:2015tya,Badger:2016ozq,Mastrolia:2016dhn,Georgoudis:2016wff,Peraro:2016wsq,Abreu:2017idw,Abreu:2017xsl}), based on ideas such as integrand reduction
and maximal cuts, are very promising, but have not reached the level
of automation yet which is provided by the tools listed in Table~\ref{tab:loopamp}.

Numerous methods for the numerical calculation of multi-loop integrals
have been developed in addition to the ones mentioned above, we list only a few more recent ones here: 
direct numerical integration in momentum space~\cite{Becker:2012bi},
dispersion relations~\cite{Bauberger:2017nct}, 
use of the loop-tree duality~\cite{Chachamis:2016olm}, 
a toolbox of various dedicated numerical techniques~\cite{Passarino:2006gv,Actis:2008ts},
numerical solution of differential
equations~\cite{Baernreuther:2013caa,Czakon:2008zk},
numerical extrapolation
method~\cite{deDoncker:2017gnb,Kato:2016vvp}, numerical evaluation of
Mellin-Barnes integrals~\cite{Czakon:2005rk,Gluza:2016fwh,Dubovyk:2016aqv},
private implementations of sector decomposition~\cite{Roth:1996pd,Denner:2004iz,Passarino:2006gv,Anastasiou:2007qb,Ueda:2009xx,Anastasiou:2010pw}.

\begin{center}
\begin{table}[h]
\caption{\label{tab:realsub}Methods for the isolation of IR divergent real
  radiation at NNLO.}
\centering
\begin{tabular}{@{}*{7}{l}}
\br
method &  analytic integr. of &type/restrictions\\
       &subtraction terms& \\
\mr
antenna subtraction~\cite{GehrmannDeRidder:2005cm}& yes&subtraction\\
$q_T$ subtraction~\cite{Catani:2009sm}&yes &slicing; colourless final states\\
N-jettiness~\cite{Boughezal:2015dva,Gaunt:2015pea}& yes&slicing &\\
sector-improved residue subtraction\cite{Heinrich:2002rc,Anastasiou:2003gr,Binoth:2004jv,Czakon:2010td,Boughezal:2011jf,Czakon:2014oma,Caola:2017dug}&no&subtraction&\\
colourful
subtraction~\cite{Somogyi:2006da,DelDuca:2016csb}&partly&subtraction;
colourless initial states\\
\br
\end{tabular}
\end{table}
\end{center}

\section{Sector decomposition}

Now we will describe the program {\sc SecDec}~\cite{Carter:2010hi,Borowka:2012yc,Borowka:2015mxa,Borowka:2017idc} in more detail. 
The sector decomposition algorithm is described in \cite{Binoth:2000ps,Heinrich:2008si},
which was inspired by earlier ideas as contained in \cite{Hepp:1966eg,Roth:1996pd}.

Higher order calculations in perturbation theory have in common 
that they involve multi-dimensional integrations over some parameters: 
Feynman (or Schwinger) parameters in the case of (multi-)loop integrals, 
or parameters related to the integration 
of subtraction terms over a factorised phase space in the case of infrared-divergent 
real radiation.
Usually, these calculations are performed within the framework of dimensional regularisation, 
and one of the challenges is to factorise the poles in the regulator $\eps$. 

The program \secdec\cite{Carter:2010hi,Borowka:2012yc,Borowka:2015mxa,Borowka:2017idc} is designed to 
perform this task in an automated way, and to integrate the 
coefficients of the resulting Laurent series in $\eps$ numerically.

The original sector decomposition algorithm described in Ref.~\cite{Binoth:2000ps} is based on 
an iterative procedure, which may run into an infinite recursion. 
It was pointed out however~\cite{Bogner:2007cr} that the structure of Feynman
integrals is such that a decomposition algorithm must exist which is guaranteed to stop, 
as the procedure can be mapped to a known problem in convex geometry.
In Ref.~\cite{Kaneko:2009qx}, an algorithm was presented which cannot lead to infinite recursion 
and is more efficient than  previously employed algorithms with this property. 
{\sc SecDec}-3 and \pysecdec{} contain the implementation of a decomposition strategy
(called $G_2$ in {\sc SecDec}-3 and {\tt geometric} in \pysecdec),  based on a modification of the method of 
Ref.~\cite{Kaneko:2009qx}, which usually outperforms the original
iterative strategy (called $X$, or {\tt iterative}). 
 
 \subsection{Feynman parameter integrals}

Multi-loop Feynman integrals can be written in a generic form.
For ease of notation, we limit ourselves to  scalar integrals here. 
Integrals with loop momenta in the numerator, or inverse propagators, 
only lead to an additional function 
in the numerator, and can be treated in the same way. 
We refer to \cite{Heinrich:2008si, Borowka:2015mxa,Schlenk:2016cwf} for further details.

A scalar Feynman integral $G$ in $D$ dimensions 
at $L$ loops with  $N$ propagators, where 
the propagators can have arbitrary, not necessarily integer powers $\nu_j$,  
has the following representation in momentum space:
\begin{eqnarray}\label{eq:integraldef}
G&=&\int\prod\limits_{l=1}^{L} \rd^D\kappa_l\;
\frac{1}
{\prod\limits_{j=1}^{N} P_{j}^{\nu_j}(\{k\},\{p\},m_j^2)}\\
\rd^D\kappa_l&=&\frac{\mu^{4-D}}{i\pi^{\frac{D}{2}}}\,\rd^D k_l\;,\;
P_j(\{k\},\{p\},m_j^2)=q_j^2-m_j^2+i\delta\;,\nn
\end{eqnarray}
where the $q_j$ are linear combinations of external momenta $p_i$ and loop momenta $k_l$.

Introducing Feynman parameters in Eq.~(\ref{eq:integraldef}) leads to
\begin{eqnarray}
G&=&  
\frac{\Gamma(N_\nu)}{\prod_{j=1}^{N}\Gamma(\nu_j)}
\int_0^\infty \,\prod\limits_{j=1}^{N}\rd x_j\,\,x_j^{\nu_j-1}\, 
\delta\big(1-\sum_{i=1}^N x_i\big)\\
&&\cdot \int \rd^D\kappa_1\ldots\rd^D\kappa_L
\left[ 
       \sum\limits_{i,j=1}^{L} k_i^{\rm{T}}\, M_{ij}\, k_j  - 
       2\sum\limits_{j=1}^{L} k_j^{\rm{T}}\cdot Q_j +J +\idel
                             \right]^{-N\nu}\nn \\
&&\nn\\
&=&\frac{(-1)^{N_{\nu}}}{\prod_{j=1}^{N}\Gamma(\nu_j)}\Gamma(N_{\nu}-LD/2)\, \int\limits_{0}^{\infty} 
\,\prod\limits_{j=1}^{N}{\rm{d}}x_j\,\,x_j^{\nu_j-1}\,\delta(1-\sum_{l=1}^N x_l)
\frac{{\cal U}^{N_{\nu}-(L+1) D/2}}
{{\cal F}^{N_\nu-L D/2}}\;,\label{eq:scalarloopint}\nn \\
 &&\nonumber\\
&&\mbox{where}  \nonumber\\
{\cal F}(\vec x) &=& \det (M) 
\left[ \sum\limits_{j,l=1}^{L} Q_j \, M^{-1}_{jl}\, Q_l
-J -\idel\right]\;,\label{DEF:F}\\
{\cal U}(\vec x) &=& \det (M),  \; \;N_\nu=\sum_{j=1}^N\nu_j\;.\label{DEF:U}
\end{eqnarray}
In the expressions above, $M$ is an $L\times L$ matrix containing Feynman parameters,
$Q$ is an $L$-dimensional vector, where each entry is a linear combination of external momenta and 
Feynman parameters, and $J$ is a scalar expression containing kinematic invariants and Feynman parameters.
%

\medskip

${\cal U}$ is a positive semi-definite function, which 
vanishes at the UV subdivergences of the graph.
In the region where all invariants formed from external momenta are
negative (``{\em Euclidean region}''), 
${\cal F}$ is also a positive semi-definite function 
of the Feynman parameters $x_j$.  If some of the invariants are zero, 
for example if some of the external momenta
are light-like, an IR divergence may appear and ${\cal F}$  vanishes for  certain 
points in parameter space. 
In the Euclidean region, 
the necessary condition ${\cal F}=0$ for an IR divergence can only 
be fulfilled if some of the parameters $x_i$ are zero.
The endpoint singularities of both UV and IR nature can be regulated by dimensional regularisation and 
factored out of the functions $ {\cal U}$ and ${\cal F}$ using sector
decomposition. 

\medskip

The basic concept of sector decomposition is the following:
We consider a two-dimensional 
parameter integral which contains a singular region where both $x$ and $y$ vanish:
\begin{eqnarray}
I&=&
\int_0^1 dx\,\int_0^1dy \,x^{-1-a\epsilon}\,y^{-b\epsilon}\,(x+y)^{-1}\;.
\end{eqnarray}
Our aim is to factorise the singularities for $x\to 0$ and  $y\to 0$. 
Therefore we divide the integration range into two 
sectors where $x$ and $y$ are ordered:
\begin{eqnarray*}
I&=&
\int_0^1 dx\,\int_0^1dy \,x^{-1-a\epsilon}\,y^{-b\epsilon}\,(x+y)^{-1}\,
[\underbrace{\Theta(x-y)}_{(1)}+\underbrace{\Theta(y-x)}_{(2)}]\;.
\end{eqnarray*}
Now we substitute  $y=x\,t$ in sector (1) and $x=y\,t$ in sector (2) 
to remap the integration range to the unit square and obtain
\begin{eqnarray}
I&=&\int_0^1 dx\,x^{-1-(a+b)\epsilon}\int_0^1 dt\,t^{-b\eps}
\,(1+t)^{-1}\nn\\
&+&\int_0^1 dy
\,y^{-1-(a+b)\epsilon}\int_0^1 dt\,t^{-1-a\epsilon}\,(1+t)^{-1}\;.
\end{eqnarray}
This way the singularities are  factorised into monomials, while the remaining denominator goes to a constant if the 
integration variables approach zero.
For more complicated integrands, this procedure can be iterated until a complete factorisation is achieved.

However, after the UV and IR  singularities have been extracted 
as poles in  1/$\epsilon$,
for non-Euclidean kinematics 
integrable singularities related to kinematic thresholds remain. 
These singularities imply that ${\cal F}$ is vanishing inside the integration region 
for some combinations of Feynman parameter values and values of the kinematic invariants.
However, the integrals can be evaluated 
by deforming the integration contour into the
complex plane~\cite{Soper:1999xk}, as explained in detail in Refs.~\cite{Borowka:2012yc,Borowka:2014aaa}.

\subsection{Program structure}

The program consists of two main parts, an algebraic and a
numerical part.
The algebraic part constructs the integrand from the list of
propagators or from the graph labels, 
performs the sector decomposition procedure to factorise the poles in the regulator $\eps$, 
the subtractions and the expansion in $\eps$, and prepares the contour
deformation in the case of non-Euclidean kinematics.
In {\sc SecDec}-3, all the algebraic steps are performed in {\tt
  Mathematica}. In the new version~\cite{Borowka:2017idc}, the algebraic
part has been completely restructured and implemented in {\tt python}, therefore
the new version is called {\pysecdec}. 

\begin{figure}[h]
\begin{minipage}{17pc}
\includegraphics[width=17pc]{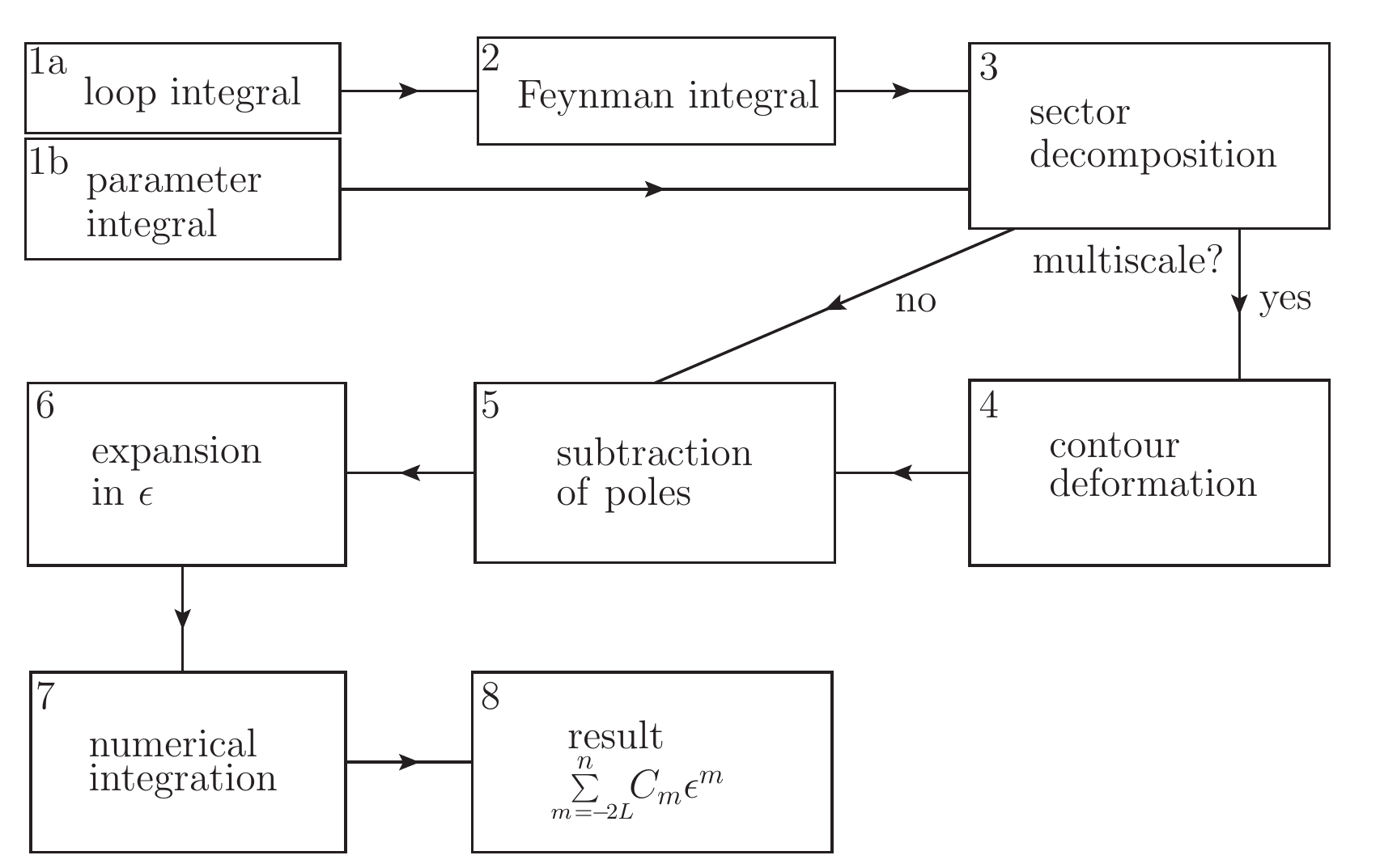}
\caption{Flowchart showing the main steps the program performs to produce the 
	numerical result as a Laurent series in $\epsilon$.
	\label{fig:structure} }
\end{minipage}
\hspace{5pc}%
\begin{minipage}{17pc}
\includegraphics[width=17pc,angle=0]{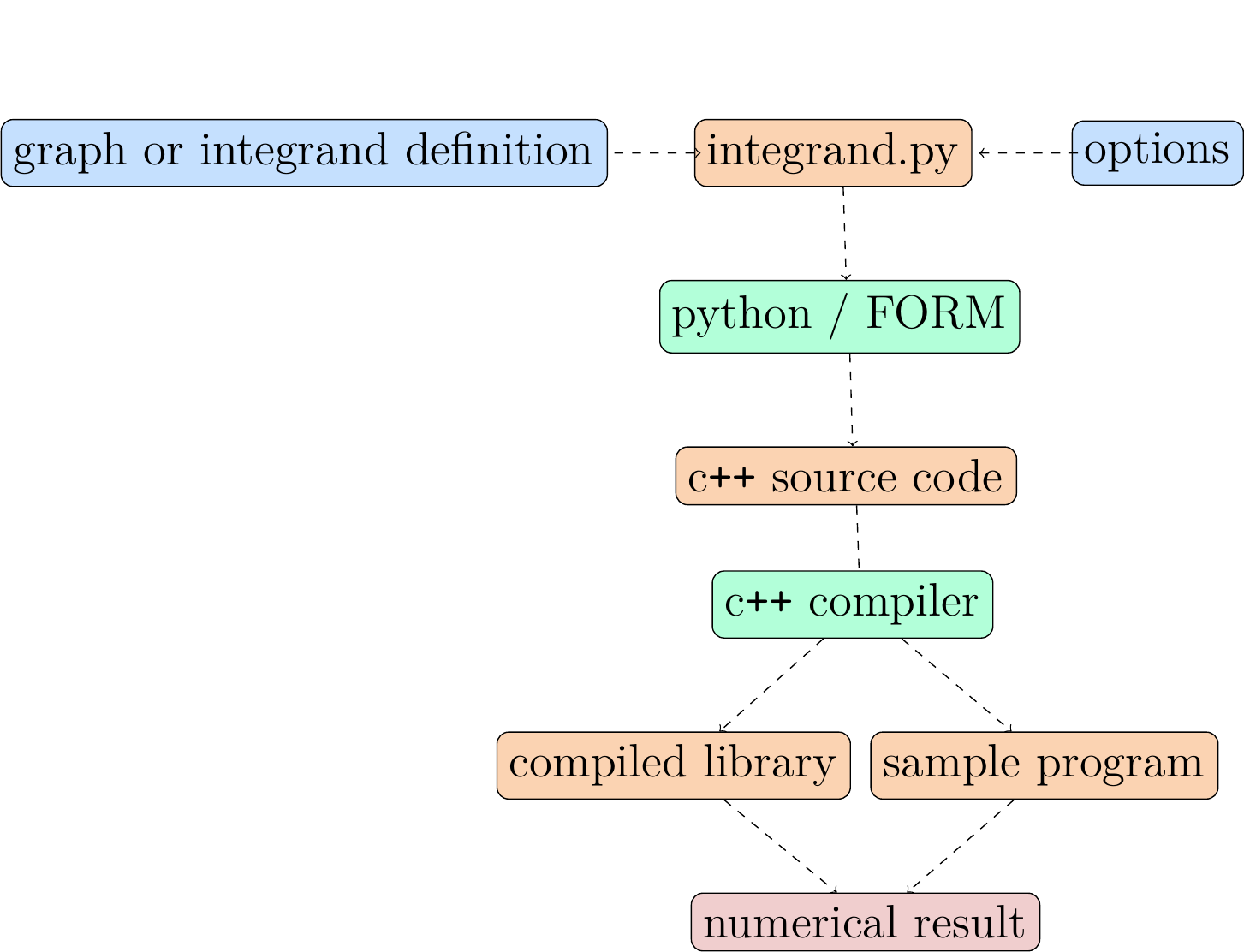}
\caption{Basic workflow of \pysecdec.\label{fig:pySD} }
\end{minipage} 
\end{figure}

The numerical part consists of C++ functions which are integrated
numerically with the {\sc Cuba} library~\cite{Hahn:2004fe}.
The new program \pysecdec{} produces C++ code using {\tt
  Form}~\cite{Vermaseren:2000nd,Kuipers:2013pba}, and in addition
produces C++ libraries such that the finite parametric functions representing an
integral after the algebraic procedure can be linked to other programs.
The basic workflow is shown in Figs.~\ref{fig:structure} and  \ref{fig:pySD}.

\subsection{Recent program developments}

In addition to the new possibilities of usage, there are various new features in \pysecdec{} compared to {\sc SecDec}-3.0: 
\begin{itemize}
\item the functions can have any number of different regulators, not
  only the dimensional regulator $\eps$, needed for example in
  analytic regularisation within Soft-Collinear Effective Theory~\cite{Becher:2011dz};
\item numerators of loop integrals can be defined in terms of
  contracted Lorentz vectors or inverse propagators or a combination
  of both;
\item the distinction between ``general functions" and ``loop integrands" is removed in the sense that all features 
which are not loop-integral-specific are also available for general polynomial functions;
\item the inclusion of ``user-defined" functions which do not enter the decomposition has been facilitated and extended; 
\item the treatment of poles which are higher than logarithmic has been improved;
\item a procedure has been implemented to detect and remap spurious singularities  which cannot be cured by contour deformation;
\item a symmetry finder 
has been added which can detect  possible isomorphisms between sectors.
\end{itemize}

Version 1 of {\sc pySecDec}~\cite{Borowka:2017idc} is
available at {\tt http://secdec.hepforge.org/}.

\subsection{Phenomenological application} 

\begin{figure}[htb]
\includegraphics[width=20pc]{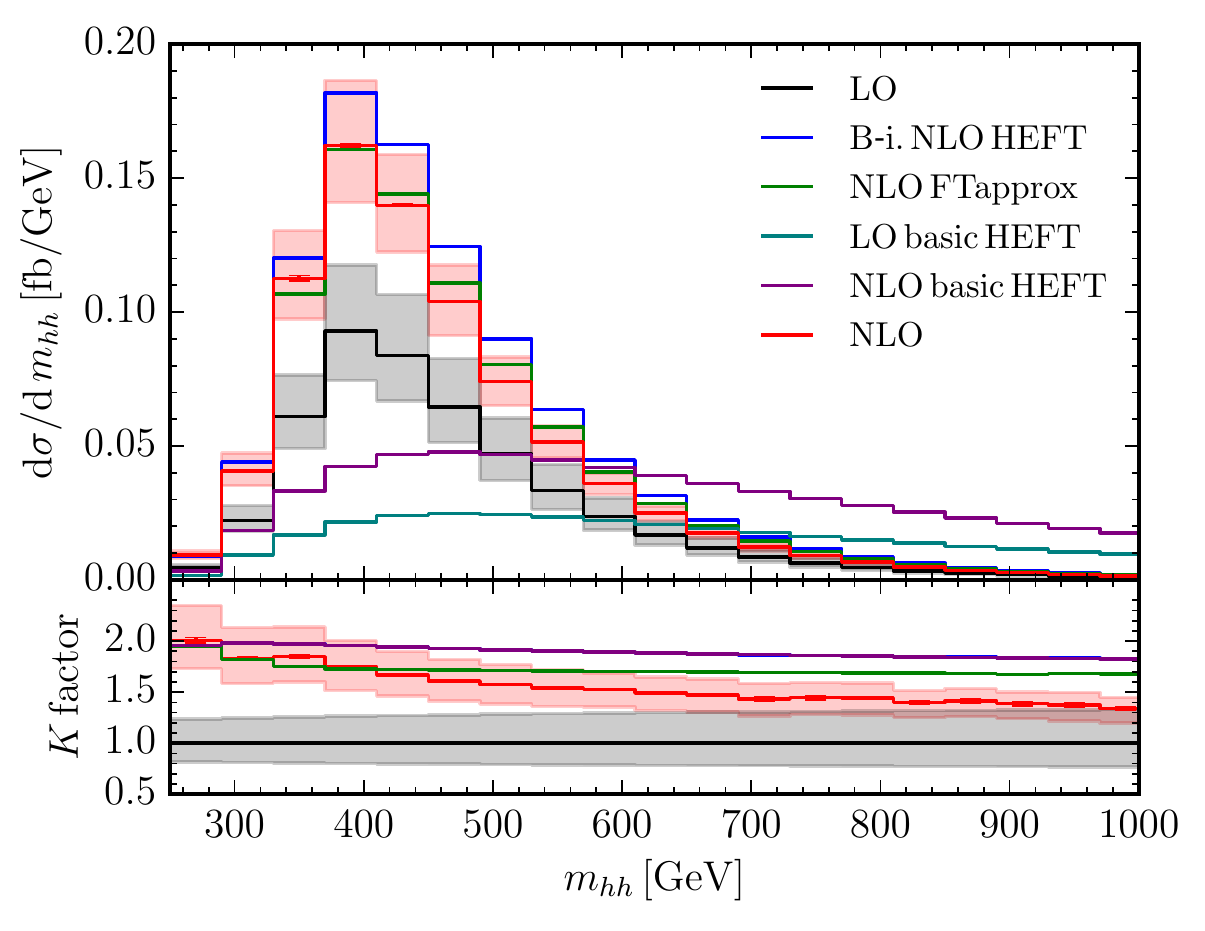}\hspace{2pc}%
\begin{minipage}[b]{18pc}\caption{Higgs boson pair invariant mass distribution with full top quark mass dependence 
compared to various approximations. B-i. NLO HEFT denotes the Born-improved HEFT approximation, while ``basic HEFT" is without the rescaling by the full Born level result. ``FTapprox" stands for an approximation where the real radiation part is calculated with full mass dependence, while the virtual part is given by the
Born-improved HEFT approximation.\label{fig:mHH}}
\end{minipage}
\end{figure}

The numerical approach based on \secdec 
has been applied to calculate
massive two-loop integrals entering $gg\to HH$ at NLO, retaining the
full top quark mass dependence~\cite{Borowka:2016ehy,Borowka:2016ypz,Heinrich:2017kxx}. 
The calculation is based on the setup described in
Refs.~\cite{Jones:2016bci,Kerner:2016msj,Borowka:2016ypz}. 
The amplitude generation leads to about 10000 integrals before any symmetries are taken into account, 
which have been reduced to ${\cal O}(300)$ integrals using {\sc Reduze}\cite{Studerus:2009ye,vonManteuffel:2012np}.
A complete reduction could not be obtained for the non-planar 4-point integrals. 
The inverse propagators appearing in unreduced integrals were
rewritten in terms of scalar products and directly computed with {\sc SecDec}.

For the total cross section at $\sqrt{s}=14$\,TeV, we found a reduction of about 14\% when including the full top quark mass dependence
as compared to the Born-improved HEFT approximation, where in the
latter the NLO corrections are calculated in the $m_t\to\infty$ limit and rescaled with the full Born level result.
Fig.~\ref{fig:mHH} shows results for the Higgs boson pair invariant mass distribution. 
For further details we refer to~\cite{Borowka:2016ehy,Borowka:2016ypz,Heinrich:2017kxx}.

\section{Conclusions}
We have given a brief overview on numerical methods to calculate integrals 
(and cross sections) beyond one-loop order, before focusing on the program {\sc SecDec}, 
in particular the new version \pysecdec.
We pointed to its application within a context that goes beyond the calculation of individual master integrals, 
for example the possibility to use it as a library to evaluate
two-loop amplitudes where the analytic expressions for the master integrals are not known.


\subsection*{Acknowledgments}
G.H. would like to thank the organizers of CPP2016 for the great
workshop, and dedicate these proceedings to the memory of
Shimizu-Sensei. 
We also would like to thank Nicolas Greiner and Tom Zirke for
collaboration on parts of the projects presented here.
This research was supported in part by the 
Research Executive Agency (REA) of the European Union under the Grant Agreement
PITN-GA2012316704 (HiggsTools) and the ERC Advanced Grant MC@NNLO
(340983). 
S. Borowka gratefully acknowledges financial support by the ERC Starting 
Grant ``MathAm" (39568).


\section*{References}
\bibliographystyle{JHEP}
\input{CPP2016proc.bbl}

\end{document}

%% file: CPP2016proc.bbl
\providecommand{\href}[2]{#2}\begingroup\raggedright\endgroup